\documentclass[reprint,aip,jcp,amsmath,amssymb]{revtex4-1}
\usepackage[utf8]{inputenc}
\usepackage{amsmath}
\usepackage{graphicx}   
\usepackage{dcolumn}    
\usepackage{bm}         
\usepackage{textcomp}
\usepackage{amssymb}
\usepackage{array,multirow}
\usepackage{microtype}
\usepackage{mathtools}

\newcommand{\CABNt}{C_{AB}^{[N]}(t)}
\newcommand{\CABMt}{C_{AB}^{[M]}(t)}
\newcommand{\eb}{e^{-\beta \hat H}}

\newcommand{\etf}{e^{-i \hat H t/\hbar}}
\newcommand{\etb}{e^{i \hat H t/\hbar}}

\newcommand{\bra}[1]{\langle #1 |}
\newcommand{\ket}[1]{| #1 \rangle}
\newcommand{\kb}[1]{\ket{#1}\bra{#1}}

\newcommand{\dd}[2]{\frac{d #1}{d #2}}

\newcommand{\ddp}[2]{\frac{\partial #1}{\partial #2}}

\newcommand{\piNz}{\prod_{i=0}^{N-1}}

\newcommand{\smiNz}{\sum_{i=0}^{N-1}}

\newcommand{\smjMM}{\sum_{j=-(M-1)/2}^{(M-1)/2}}
\newcommand{\smkMM}{\sum_{k=-(M-1)/2}^{(M-1)/2}}

\newcommand{\pjMM}{\prod_{j=-(M-1)/2}^{(M-1)/2}}

\newcommand{\ltti}{\lim_{t\to \infty}}

\newcommand{\lNti}{\lim_{N\to \infty}}

\newcommand{\largeN}{$N\to\infty$}

\newcommand{\tr}{ {\rm Tr} }

\newcommand{\no}{\nonumber}

\newcommand{\betaN}{\beta_N}
\newcommand{\ola}{\overleftarrow}
\newcommand{\ora}{\overrightarrow}

\newcommand{\bp}{ {\bf p} }

\newcommand{\bq}{ {\bf q} }

\newcommand{\bz}{ {\bf z} }
\newcommand{\bDelta}{ {\bf \Delta} }

\newcommand{\bQ}{ {\bf Q} }

\newcommand{\bP}{ {\bf P} }

\newcommand{\inti}{\int_{-\infty}^{\infty}}
\newcommand{\eqr}[1]{Eq.~\eqref{eq:#1}}
\newcommand{\eqsr}[2]{Eqs.~\eqref{eq:#1} and \eqref{eq:#2}}
\newcommand{\eql}[1]{\label{eq:#1}}
\newcommand{\figr}[1]{Fig.~\ref{fig:#1}}
\newcommand{\figl}[1]{\label{fig:#1}}

\newcommand{\LMat}{\mathcal{L}_{\rm Mat}^{[M]}}

\newcommand{\LIm}{\mathcal{L}_{\Im}^{[M]}}

\newcommand{\LmRP}{\mathcal{L}_{\rm RP}^{[M]}}

\newcommand{\UMQ}{U^{[M]}(\bQ)}

\begin{document}

\title{How to obtain thermostatted ring polymer molecular dynamics from exact quantum dynamics and when to use it} 
\author{Timothy J.~H.~Hele}
\thanks{Corresponding author: tjhh2@cam.ac.uk}
\affiliation{Department of Chemistry, University of Cambridge, Lensfield Road, Cambridge, CB2 1EW, UK.}
\date{\today}

\begin{abstract}
We obtain thermostatted ring polymer molecular dynamics (TRPMD) from exact quantum dynamics via Matsubara dynamics, a recently-derived form of linearization which conserves the quantum Boltzmann distribution. Performing a contour integral in the complex quantum Boltzmann distribution of Matsubara dynamics, replacement of the imaginary Liouvillian which results with a Fokker-Planck term gives TRPMD. We thereby provide error terms between TRPMD and quantum dynamics and predict the systems in which they are likely to be small. Using a harmonic analysis we show that careful addition of friction causes the correct oscillation frequency of the higher ring-polymer normal modes in a harmonic well, which we illustrate with calculation of the position-squared autocorrelation function. 
However, no physical friction parameter will produce the correct fluctuation dynamics for a parabolic barrier.
The results in this paper are consistent with previous numerical studies and advise the use of TRPMD for the computation of spectra.
\emph{This manuscript has been submitted to Molecular Physics. If accepted for publication, it will be available at http://wwww.tandfonline.com}
\end{abstract}


\maketitle

\section{Introduction}
The computation of thermal time-correlation functions is of central importance in chemical physics \cite{gar09,zwa01} in order to evaluate many physically observable quantities such as reaction rates, diffusion constants, spectra and scattering data. \cite{hab13,liu15} 

%
Exact evaluation of the quantum correlation function scales exponentially with system size and so is impractical for more than a few atoms. \cite{hab13} There is consequently a need for computationally tractable approximations to quantum time-correlation functions \cite{liu15,mil05rev}, preferably which are known to be equivalent to the quantum result in certain limits, and for which the likely error is known in advance of calculation. One crude solution is to use a purely classical correlation function, which will scale linearly with system size. However, the classical Boltzmann distribution is highly inaccurate for many systems such as water at room temperature \cite{hab09}, and ignores effects such as tunnelling and zero-point energy. \cite{hab13,mil05rev} There is consequently a need to incorporate quantum statistics into such calculations, but with approximate, preferably classical-like, dynamics.

Various approaches have been developed, including the ``classical Wigner'' or linearized semiclassical initial value representation (LSC-IVR) method \cite{wig32qm,liu15,liu07,shi03} which truncates the (exact) Moyal series \cite{moy49} for time evolution at $\hbar^0$, Centroid Molecular Dynamics (CMD) \cite{cao94_1,cao94_2,cao94_3,cao94,cao94_5,vot96,kin98,rei00}, which propagates the path-integral centroid in the mean-field of the other path-integral normal modes, and Ring Polymer Molecular Dynamics (RPMD) \cite{man04,man05che,man05ref,hab13} which takes the classical dynamics of a ring polymer \cite{par84} literally. 

All these methods have various limitations; LSC-IVR does not conserve the quantum Boltzmann distribution leading to zero-point energy leakage \cite{hab09,liu15}, whereas CMD and RPMD both fail for multidimensional spectra \cite{wit09,iva10}; CMD has the \emph{curvature problem} where peaks are broadened and red-shifted whereas RPMD has \emph{spurious resonances} where the ring polymer springs couple to frequencies in the potential leading to splitting of the physical peak. \cite{ros14} 

Recently, Thermostatted Ring Polymer Molecular Dynamics (TRPMD) has been introduced, which applied a Langevin thermostat \cite{cer10phd,nit06,bus07} to the dynamics of the ring polymer \cite{cer10,ros14,ros14com}. This was originally conceived for the evalutation of static properties \cite{cer10}, but it appeared to be remarkably successful for the computation of spectra \cite{ros14,ros14com}, accurately replicating multidimensional spectra where CMD and RPMD fail and correctly predicting the diffusion and rotational constants of liquid water. Like RPMD, the short-time, transition-state theory (TST) limit of the TRPMD flux-side correlation function is identical to quantum transition-state theory (QTST) \cite{ros14,hel13,alt13,hel13unique,hel14phd}: the instantaneous thermal ring-polymer flux through a position-space dividing surface is equal to the intantaneous thermal quantum flux, and the TRPMD rate will equal the exact quantum rate in the absence of recrossing by either the quantum dynamics or TRPMD dynamics \cite{alt13}. Because TRPMD obeys detailed balance, its reaction rate is independent of the location of the dividing surface \cite{hel15trpmd}, as is the case for RPMD and CMD but not many TST-based methods.

Nevertheless, TRPMD is not without its faults; like RPMD and CMD it fails to capture effects such as a Fermi resonance involving a fourth-order coupling in the Zundel cation \cite{ros14}, and beneath the crossover temperature (see \eqr{xover}) application of friction to reaction rates causes them to decrease, resulting in a less accurate result compared to RPMD for symmetric systems, and a more accurate result (but with an adjustable parameter whose value is not determinable in advance) for asymmetric systems. \cite{hel15trpmd}


Very recently, both RPMD and CMD have been obtained from the exact quantum time-correlation function via ``Matsubara dynamics'', a form of linearization which conserves the quantum Boltzmann distribution \cite{hel15,hel15rel}. Matsubara dynamics results from discarding fluctuations of the very high frequency path-integral normal modes (higher frequencies than those required to converge the quantum Boltzmann distribution) from the (exact) Moyal bracket and is inherently classical as well as satisfying detailed balance \cite{hel15}. However, Matsubara dynamics is not amenable to computation in large systems since it suffers from the sign problem due to a phase factor in the complex quantum Boltzmann distribution \cite{hel15rel}. An approximation to Matsubara dynamics where the centroid moves in the mean field of the other Matsubara modes leads to CMD, and by moving the momentum contour in the quantum Boltzmann distribution and discarding the imaginary Liouvillian which results, RPMD arises \cite{hel15rel}.

Obtaining RPMD and CMD from the exact quantum expression provides analytical expressions for their error from the quantum result, such that it can be known \emph{a priori} whether they will function well in a given system, whereas previously one had to rely on induction from earlier numerical studies on systems for which RPMD or CMD had been successful, though there was no guarantee that such reasoning would extend to a new system. This was seen, for example, in RPMD rate theory failing in the Marcus inverted regime \cite{men11} despite being very successful for rate computation in a large variety of other systems \cite{man05che,man05ref,col09,col10,sul11}. With the derivation of QTST \cite{hel13,alt13,hel13unique,hel14phd}, it can be known \emph{a priori} that a system whose optimal ring-polymer dividing surface will be significantly recrossed by the quantum dynamics will not have its rate accurately computed by RPMD \cite{hel14phd}.

Consequently, investigating whether TRPMD could also be obtained from exact quantum time evolution and thereby discerning the situations where it is likely to work \emph{a priori}, rather than relying on the (small but growing) literature of its application to physical systems \cite{ros14,ros14com,hel15trpmd}, would be of considerable benefit to the field.


In this paper we obtain TRPMD from exact quantum dynamics by showing that it is a stochastic approximation to Matsubara dynamics. To obtain a computationally tractable approximation to Matsubara dynamics, we move the momentum contour in the complex plane in order to convert the complex quantum Boltzmann distribution into the real ring polymer distribution. This transformation generates a complex Liouvillian in the dynamics \cite{hel15rel}, which is not in itself amenable to computation due to the complex trajectories which result. Previously, the imaginary part of the Liouvillian was simply discarded, shifting the oscillation frequencies in the higher normal modes and leading to RPMD, but here we replace it with a Fokker-Planck term, producing TRPMD. 

To examine the effect of the friction matrix we conisder a harmonic well and a parabolic barrier, for which the correlation function (if defined) can be evaluated analytically. We find that a unique and system-independent value of the friction matrix causes all normal modes to oscillate at the correct (external) frequency in a harmonic well, and illustrate this with the position-squared autocorrelation function, where TRPMD has the correct zero-time value and frequency; neither RPMD nor CMD can reproduce both these properties \cite{hor05,jan14}. We then examine a parabolic barrier, where CMD and RPMD have the incorrect fluctuation dynamics; the higher normal modes in RPMD being bound (above the relevant crossover temperature, see \eqr{xover}), rather than unbound as in Matsubara dynamics. Here application of any meaningful (i.e.\ positive) friction does not cause the erroneously bound normal modes in TRPMD to become scattering, and nor does it cause unbound modes to have the correct escape frequency, meaning that application of friction is unlikely to assist in the accuracy of reaction rate or diffusion calculation. \cite{hel15trpmd}


We begin by revisiting Matsubara dynamics in section~\ref{sec:mat}, followed by obtaining TRPMD in section~\ref{sec:trpmd} and examining the friction matrix in section~\ref{sec:fric}, before presenting conclusions in section~\ref{sec:con}.

\section{Summary of Matsubara dynamics}
\label{sec:mat}
For simplicity, we consider a one-dimensional system with mass $m$, co-ordinate $q$, and Hamiltonian $\hat H = \hat p^2/2m + V(\hat q)$, where $V(q)$ is the potential.\footnote{Here we consider dynamics on a single Born-Oppenheimer potential energy surface and at temperatures sufficiently high that Bose-Einstein and Fermi-Dirac statistics need not be considered, which is the case for most systems to which CMD, RPMD and TRPMD have been applied.} Extensions to further dimensions follows immediately and merely requires more indices. The Kubo-transformed thermal quantum time-correlation function at inverse temperature $\beta \equiv 1/k_{\rm B}T$ is \cite{kub57}
\begin{align}
 c_{AB}(t) = \frac{1}{\beta}\int_0^{\beta} d\sigma \tr \left[e^{-\sigma \hat H} \hat A e^{-(\beta-\sigma) \hat H} \etb \hat B \etf \right] \eql{kub}
\end{align}
and which can easily be related to the conventional asymmetric-split quantum correlation function \cite{man04,liu15}. If $\hat A$ and $\hat B$ are linear operators in position or momentum, \eqr{kub} is identical to the Generalized Kubo transform \cite{hel13,alt13,hel13unique}
\begin{align}
 \CABNt = & \int d\bq \int d\bz \int d\bDelta \ A(\bq) B(\bz) \no\\
 & \times \piNz \bra{q_{i-1} - \Delta_{i-1}/2} \eb \ket{q_{i} + \Delta_i/2}\no\\
 & \times \bra{q_{i} + \Delta_i/2} \etb \kb{z_i} \etf \ket{q_i - \Delta_i/2} \eql{genk}
\end{align}
where $\int d\bq = \piNz \inti dq_i$ (similarly for $\int d\bz$ and $\int d\bDelta$),
\begin{align}
 A(\bq) = \frac{1}{N}\smiNz A(q_i)
\end{align}
and likewise for $B(\bq)$. 

The full derivation of Matsubara dynamics is given in Ref.~\onlinecite{hel15} and here we sketch the relevant details. 
To obtain the time-evolution of \eqr{genk} in the phase-space representation, we take its Wigner Transform \cite{wig32qm} and differentiate w.r.t.\ time to obtain a Moyal series \cite{moy49, hel15} which is formally exact. 
We then transform the correlation function from bead co-ordinates $(\bp,\bq)$ to ring-polymer normal mode co-ordinates \cite{ric09} $(\bP,\bQ)$, as detailed in appendix~\ref{ap:mat}, such that $Q_0$ and $P_0$ are the position and momentum centroids respectively.

Truncating the resulting Moyal series (either in the normal mode or bead representation) at $\mathcal{O}(\hbar^0)$ leads to the linearized semiclassical initial value representation (LSC-IVR) \cite{wan98,sun98,liu15,shi03}, which involves propagating trajectories under the classical Hamiltonian of the system drawn from a Wigner-transformed quantum Boltzmann distribution. Conversely, truncating the Moyal bracket to the lowest $M$ `Matsubara' normal modes (see appendix~\ref{ap:mat}) results in a dynamics which is inherently classical (all powers of $\mathcal{O}(\hbar^2)$ and higher vanish from the Moyal series without further approximation) and which conserves the quantum Boltzmann distribution and satisfies detailed balance \cite{hel15}, unlike LSC-IVR \cite{liu15,hab09}. This leads to a classical-like correlation function \cite{hel15}
\begin{align}
 \CABMt = & \frac{\alpha_M}{2\pi\hbar} \int d\bP \int d\bQ \ e^{-\beta [H_M(\bP,\bQ) -i \theta_M(\bP,\bQ)]} \no\\
 & \qquad \times A(\bQ) e^{\LMat t} B(\bQ) \eql{mat}
\end{align}
where the Matsubara Hamiltonian is 
\begin{align}
 H_M(\bP,\bQ) = \smjMM \frac{P_j^2}{2m} + \UMQ,
\end{align}
$\UMQ$ is defined in the appendix, $\alpha_M = \hbar^{M-1}[(M-1)/2]!^2$, and the phase factor is
\begin{align}
  \theta_M(\bP,\bQ) = \smjMM P_j \tilde \omega_j Q_{-j} 
\end{align}
where $\tilde \omega_j = 2\pi j/\beta \hbar$ are the Matsubara frequencies \cite{mat55} which, in this definition, can be negative. The integrals are taken to mean $\int d\bP = \pjMM \inti d P_j$ and likewise for $\int d\bQ$. Matsubara dynamics is defined by the Liouvillian
\begin{align}
 \LMat= &  \frac{\bP}{m} \ora\nabla_{\bQ} - \UMQ \ola\nabla_{\bQ} \cdot \ora \nabla_{\bP} \eql{lmat} 
\end{align}
such that $\LMat \equiv \{\cdot,H_M(\bP,\bQ)\}$ where $\{\cdot,\cdot\}$ is the classical Poisson bracket. \cite{nit06}

\section{Emergence of TRPMD}
\label{sec:trpmd}
The Matsubara correlation function in \eqr{mat} suffers from the sign problem, such that it is not amenable to computation in complex systems. 
To make the distribution real, we continue into the complex plane of $\bP$ with
\begin{align}
 \bar P_{j} = P_{j} - i m\tilde \omega_j Q_{- j} \eql{momtran}
\end{align}
for all $j$ (such that no analytic continuation is necessary for the momentum centroid) to give
\begin{align}
  \mathcal{L}^{[M]}_{\bar \bP} = & \mathcal{L}^{[M]}_{\rm RP} + i \mathcal{L}^{[M]}_{\Im} 
\end{align}
where $\mathcal{L}^{[M]}_{\rm RP}$ is the ring polymer Liouvillian,
\begin{align}
 \mathcal{L}^{[M]}_{\rm RP} 
 = \smjMM \frac{\bar P_j}{m} \ddp{}{Q_j} - \left[\ddp{\UMQ}{Q_j} + m \tilde\omega_j^2 Q_j \right] \ddp{}{\bar P_j}
\end{align}
and the imaginary component of the Liouvillian is
\begin{align}
 \mathcal{L}^{[M]}_{\Im} = \smjMM \tilde\omega_j \left( \bar P_j \ddp{}{\bar P_{-j}} - Q_j \ddp{}{Q_{-j}} \right).
\end{align}
This transformation also converts the complex Matsubara distribution into the real ring polymer distribution,
\begin{align}
 e^{-\beta [H_M(\bP,\bQ) -i \theta_M(\bP,\bQ)]} = e^{-\beta R_M(\bar \bP,\bQ)}
\end{align}
where the ring-polymer Hamiltonian is
\begin{align}
 R_M(\bar \bP,\bQ) = \smjMM \left(\frac{\bar P_j^2}{2m} + \tfrac{1}{2}m\tilde \omega_j^2 Q_j^2\right) + \UMQ.
\end{align}
Both $\LmRP$ and $\LIm$ independently conserve the quantum Boltzmann distribution. 

In Appendix~\ref{ap:contour}, we prove that the complex dynamics generated by $\mathcal{L}^{[M]}_{\bar \bP}$ is analytic everywhere in the complex plane, and by contour integration of \eqr{mat} it rigorously follows
\begin{align}
 \CABMt = & \frac{\alpha_M}{2\pi\hbar} \int d\bar \bP \int d\bQ  \ e^{-\beta R_M(\bar \bP,\bQ)} \no\\
 & \qquad\qquad \times A(\bQ) e^{(\LmRP + i\LIm) t} B(\bQ) \no\\
 & + \mathcal{E}(t) \eql{contm} 
 \end{align}
where $\mathcal{E}(t)$ corresponds to the vertical edges of the integration contour; in Appendix~\ref{ap:contour} we give evidence to show that in many cases the edge term will vanish, though for an arbitrary system propagated to a finite time it is, strictly speaking, part of the error term between Matsubara dynamics and RPMD/TRPMD.

Although the real ring-polymer distribution in \eqr{contm} would, \emph{prima facie}, allow evaluation of the correlation function by inexpensive Monte Carlo techniques, the presence of $i\LIm$ in \eqr{contm} causes unstable trajectories to emerge \cite{ben08, aar10trust} which are no easier to treat numerically than the sign problem in the complex Matsubara distribution. \cite{hel15rel}

In previous research \cite{hel15rel} it was shown that approximating \eqr{contm} by discarding $\LIm$, in order to make the trajectories real but still conserve the quantum Boltzmann distribution, produces RPMD. This approximation raises the oscillation frequency of the higher ($j\neq 0$) normal modes; in a harmonic potential with external frequency $\omega_h$ they oscillate at $\bar \omega_{j} = \sqrt{\omega_h^2 + \tilde \omega_j^2}$. This is the origin of the `spurious resonances' of RPMD in multidimensional spectra \cite{iva10,wit09} and the qualitative failure of RPMD at calculating the position-squared autocorrelation function \cite{hor05,jan14}. It was also shown that a mean-field approximation to \eqr{contm}, where the centroid moves in the mean field of the other ring polymer modes, leads to CMD \cite{hel15}.

This naturally motivates investigating whether there is some other approximation to the dynamics in \eqr{contm} which, like RPMD, is real and conserves the quantum Boltzmann distribution, but unlike RPMD has the correct oscillation frequencies of the higher normal modes\footnote{The higher normal modes are not explicitly represented in CMD, though are sometimes used as a computational device to construct the mean-field potential{\protect \cite{hon06}}.}, and possibly also has the correct fluctuation dynamics at barriers. Addition of a friction (Langevin) term to the dynamics of a harmonic oscillator is known to decrease the oscillation frequency \cite{nit06} and we therefore define a stochastic dynamics by the Fokker-Planck adjoint operator
\begin{align}
 \mathcal{A}_{\rm RP}^{[M]\dag} = & \mathcal{L}^{[M]}_{\rm RP} + \mathcal{A}_{\rm wn}^{[M]\dag}
\end{align}
where the white-noise thermostat which conserves the ring-polymer distribution $e^{-\beta R_M(\bar \bP,\bQ)}$ is
\begin{align}
 \mathcal{A}_{\rm wn}^{[M]\dag} = - \bar\bP \cdot \bm{\Gamma} \cdot \nabla_{\bar\bP} + \frac{m}{\beta}\nabla_{\bar\bP} \cdot \bm{\Gamma} \cdot \nabla_{\bar\bP}
\end{align}
with $\bm{\Gamma}$ an $M\times M$ positive semidefinite friction matrix. This allows us to approximate \eqr{contm} as
\begin{align}
 \CABMt \simeq & \frac{\alpha_M}{2\pi\hbar} \int d\bar \bP \int d\bQ \ e^{-\beta R_M(\bar\bP,\bQ)} A(\bQ)  e^{\mathcal{A}_{\rm RP}^{[M]\dag}t} B(\bQ), \eql{trpmd}
\end{align}
which is TRPMD.\footnote{Strictly speaking, this is TRPMD with Matsubara rather than ring-polymer frequencies, but will converge to conventional TRPMD in the limit of large $M${\protect \cite{hel15rel}}} The error term between the quantum result and TRPMD is therefore discarding the dynamics of the highest $(N-M)$ normal modes to give Matsubara dynamics (see Eq.~(B2) of Ref.~\onlinecite{hel15}), the edges of the contour used in analytic continuation (which we suspect to be zero, see \eqr{edge}), and the difference between the TRPMD and Matsubara propagators, namely $i\LIm - \mathcal{A}_{\rm wn}^{[M]\dag}$.

\section{Friction considerations}
\label{sec:fric}
There are already numerical studies of the effect of friction on various quantities computed by TRPMD \cite{ros14,hel15trpmd}, and here we take a more theoretical approach in light of obtaining TRPMD from quantum dynamics in the previous section. Since TRPMD is an approximation to Matsubara dynamics, we seek to determine an optimal friction parameter to reduce or remove the `side-effects' of the analytic continutation to form RPMD, namely the incorrect frequencies of the higher normal modes in a bound system \cite{hab13}, and the incorrect fluctuation dynamics in an unbound (scattering) system \cite{hel15rel}.

\subsection{Harmonic well}
\label{ssec:hw}
For a model bound system, we study the harmonic potential
\begin{align}
 V(q) = \tfrac{1}{2} m\omega_h^2q^2 \eql{har}
\end{align}
for which the ring polymer normal modes decouple and the dynamics can be solved exactly, and we detemine which elements of a diagonal friction matrix will cause oscillation at a correct (external) frequency $\omega_h$, as is the case for analytically continued Matsubara dynamics [\eqr{contm}] in a harmonic potential, \cite{hel15rel}
\begin{align}
  Q_j(t) = Q_j\cos(\omega_h t) + \frac{\bar P_j}{m\omega_h}\sin(\omega_h t) + i\frac{\tilde \omega_j}{\omega_h}Q_{-j} \sin(\omega_h t). 
\end{align}

For TRPMD, the trajectories are not deterministic and we define the time-evolved phase-space density $\mathcal{Q}_j(t) \equiv \mathcal{Q}_j(Q_j,P_j,t)$ which is evolved with $\mathcal{A}_{\rm RP}^{[M]\dag}$ from initial conditions of $(Q_j,P_j)$ at $t=0$. For a harmonic potential and moderate (underdamped) friction, the time-evolution of $\mathcal{Q}_j(t)$ can be solved analytically as \cite{zwa01,nit06}
\begin{align}
 \mathcal{Q}_{j}(t) = e^{-\Gamma_{jj}t/2} \Bigg[ & Q_j\cos(\acute\omega_j t) \no\\
 & + \left(\frac{P_j}{m\acute\omega_j} + \frac{Q_j\Gamma_{jj}}{2\acute\omega_j}\right)\sin(\acute\omega_j t)\Bigg] \eql{well}
\end{align}
where the observed (damped) frequency of oscillation is
\begin{align}
 \acute\omega_j = \sqrt{\omega_h^2 + \tilde \omega_j^2 - \Gamma_{jj}^2/4}. \eql{coj}
\end{align}
%
We immediately see $\Gamma_{jj'} = 2|\tilde \omega_j| \delta_{jj'}$ will ensure that $\acute \omega_j = \omega_h$ and oscillation at the correct external frequency, a result previously suggested on the grounds of minimizing the Hamiltonian correlation time for a ring polymer in a harmonic potential, and thereby optimizing statistical sampling \cite{cer10, ros14}. To investigate different friction strengths related to this we therefore define a parameter $\lambda$ such $\Gamma_{jj'} = 2\lambda |\tilde \omega_j| \delta_{jj'}$.

Although the position-squared and momentum-squared correlation functions will oscillate at the external frequency with $\lambda=1$ (see below), examination of the position-position spectrum for a given normal mode \cite{nit06,cer10phd}
\begin{align}
 C_{Q_jQ_j}^{\rm TRPMD}(\omega) \propto \frac{1}{(\omega_h^2 + \tilde\omega_j^2 - \omega^2) + \gamma^2\omega^2}
\end{align}
shows that the maximum in the spectrum will be at $\omega = \sqrt{\omega_h^2 + \tilde \omega_j^2 - \Gamma_{jj}^2/2}$, suggesting a friction parameter of $\lambda = 2^{-1/2}$. Furthermore, consideration of the momentum spectrum $C_{P_jP_j}^{\rm TRPMD}(\omega)=m^2\omega^2 C_{Q_jQ_j}^{\rm TRPMD}(\omega)$ shows that the maximum in the momentum spectrum is always at the (erroneously high) ring polymer frequency $\omega = \sqrt{\omega_h^2 + \tilde \omega_j^2}$ and increasing friction merely broadens the peak.

\subsection{Numerical example}

\begin{figure}[tb]
 \includegraphics[width=\columnwidth]{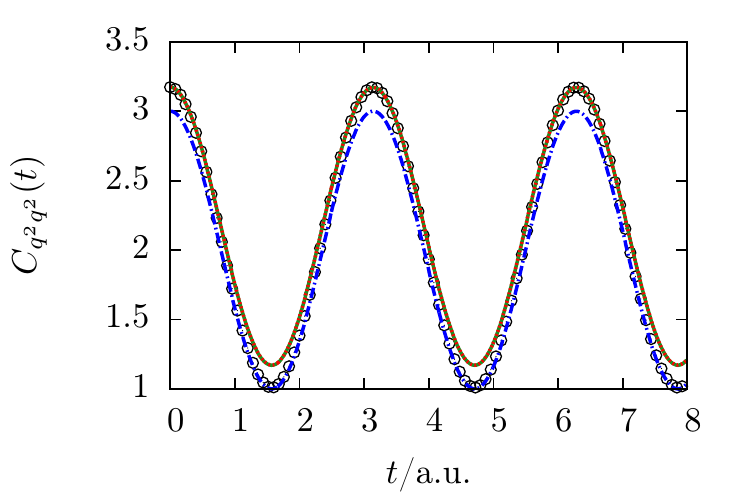}
 \includegraphics[width=\columnwidth]{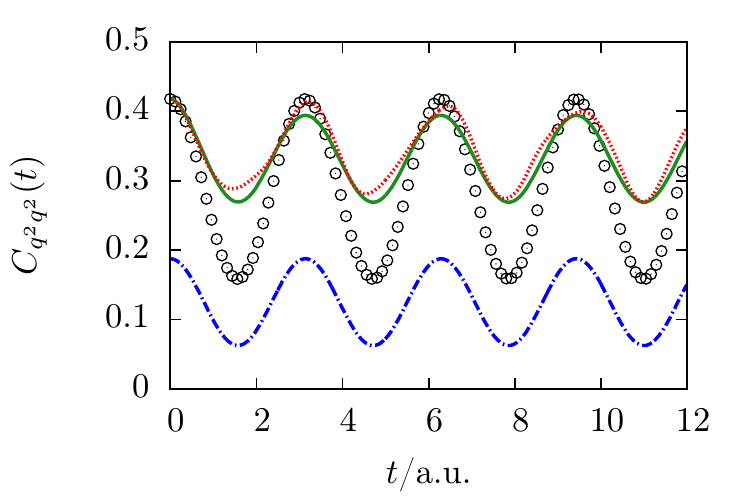}
 \includegraphics[width=\columnwidth]{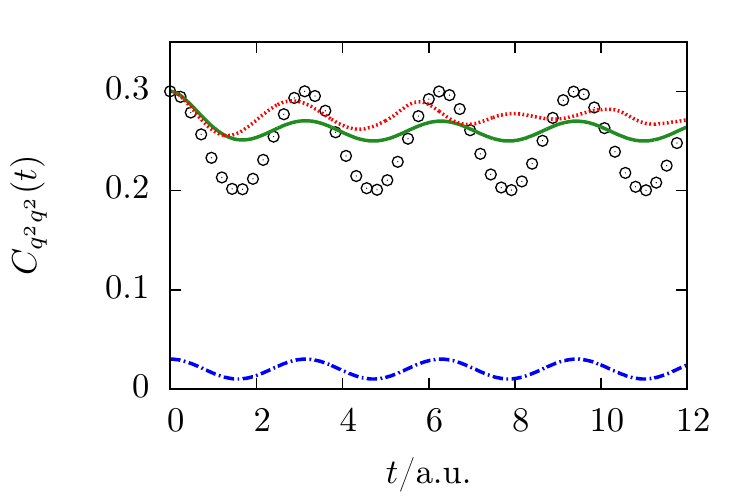}
 \caption{Position-squared autocorrelation function for a harmonic oscillator, with $\beta = 1$ (top), $\beta = 4$ (middle) and $\beta = 10$ (bottom). Black circles, quantum; solid green line, TRPMD (with optimal damping); red dots, RPMD; blue dot-dashes, CMD.}
 \figl{qq}
\end{figure}

CMD, RPMD and TRPMD correlation functions and spectra have already been the subject of many numerical studies \cite{kin98,hab13,vot96,ros14,ros14com,hel15trpmd,kre13,men11,wit09,iva10,hab09,sul11,sul14} whose results have been broadly summarized in the introduction. To clarify the nature of the approximations inherent in CMD, RPMD and TRPMD from Matsubara dynamics, we examine the position-squared autocorrelation function for a harmonic oscillator, for which Matsubara dynamics is exactly equal to the quantum result but both RPMD and CMD fail to qualitatively reproduce \cite{hor05}. CMD produces the incorrect result at $t=0$ but then oscillates at the correct frequency (though incorrect amplitude), whereas RPMD is exact at zero time but then deviates wildly from the quantum result at finite time due to the presence of the spurious frequencies in the higher normal modes \cite{hor05,jan14}. The ``nonlinear operator'' problem for which this is the archetypal model not only occurs in toy systems but is also observed in inelastic neutron scattering \cite{cra06,hab13}.

The exact quantum position-squared autocorrelation function in the harmonic potential \eqr{har} is \cite{hor05,jan14}
\begin{align}
  c_{q^2q^2}(t) = \frac{\hbar^2}{4m^2 \omega_h^2} \Bigg[&\frac{2}{\beta\hbar\omega_h}\coth\left(\frac{\beta\hbar\omega_h}{2}\right) \cos(2\omega_h t)\no\\
  & + 2 \coth^2\left(\frac{\beta\hbar\omega_h}{2}\right) - 1 \Bigg]. \eql{qqqm}
\end{align}
which in appendix~\ref{ap:qm} we show is exactly replicated by the Matsubara correlation function.
For RPMD, it is \cite{hor05,jan14}\footnote{The RPMD and TRPMD correlation functions given here use the Matsubara frequencies $\tilde \omega_j$, and converge to the conventional form using the ring-polymer frequencies $\omega_j$ in {\protect\eqr{rpfreq}} in the large $M$ and large $N$ limit. The numerical results use the ring-polymer frequencies with $N = 501$, and their convergence with the same correlation function computed with Matsubara frequencies ($M=501$) was checked.}
\begin{widetext}
\begin{align}
 C_{q^2q^2}^{\rm RPMD}(t) = \frac{1}{\beta^2m^2}\smjMM \frac{1}{\omega_h^2 + \tilde \omega_j^2} \left\{\frac{2 \cos^2[(\omega_h^2 + \tilde\omega_j^2)^{1/2} t]}{\omega_h^2 + \tilde \omega_j^2} 
 + \smkMM \frac{1}{\omega_h^2 + \tilde \omega_k^2}\right\}
\end{align}
whereas the TRPMD result for the optimal damping frequencies $\Gamma_{jj'} = 2|\tilde \omega_j|\delta_{jj'}$ is
\begin{align}
  C_{q^2q^2}^{\rm TRPMD}(t) = & \frac{1}{\beta^2m^2}\smjMM \frac{1}{\omega_h^2 + \tilde \omega_j^2} \Bigg\{\frac{2 e^{- 2|\tilde \omega_j| t}}{\omega_h^2 + \tilde \omega_j^2} 
 \left[\cos(\omega_h t) + \frac{\tilde \omega_j}{\omega_h} \sin(\omega_h t)\right]^2 + \smkMM \frac{1}{\omega_h^2 + \tilde \omega_k^2}\Bigg\}.
\end{align}
\end{widetext}

For comparison, the CMD position-squared autocorrelation function (using the CMD with classical operators method \cite{hor05,vot96,ros14,kin98}\footnote{We note that there are many other approaches of varying mathematical complexity and accuracy for the computation of general correlation functions with CMD{\protect \cite{cao94_1,cao94_2,cao94_3,cao94,vot96,hor05,rei00}}, and here restrict ourselves to methods which simply require direct computation of a correlation function.}) is
\begin{align}
 C_{q^2q^2}^{\rm CMD}(t) = \frac{1}{(\beta m \omega_h^2)^2} \left[2\cos(\omega_h t)^2 + 1 \right].
\end{align}
We use parameters to facilitate comparison with previous literature \cite{hor05}; $\hbar=k_{\rm B} = m=\omega_h=1$ and results for systems of varying $\beta$ are presented in \figr{qq}.

At high temperatures ($\beta = 1$), all methods are a good approximation to the quantum result and the RPMD and TRPMD results are indistinguishable to within graphical accuracy. At $\beta = 4$, the amplitude of oscillations is incorrect for all methods, though TRPMD starts at the correct value whereas CMD is too low. The RPMD correlation function shows deviations from harmonic behaviour due to the higher normal modes. At $\beta = 10$, the CMD correlation function is far too small and RPMD cannot replicate the oscillations, unlike TRPMD.

\begin{figure*}[tbh]
 \includegraphics[width=0.8\textwidth]{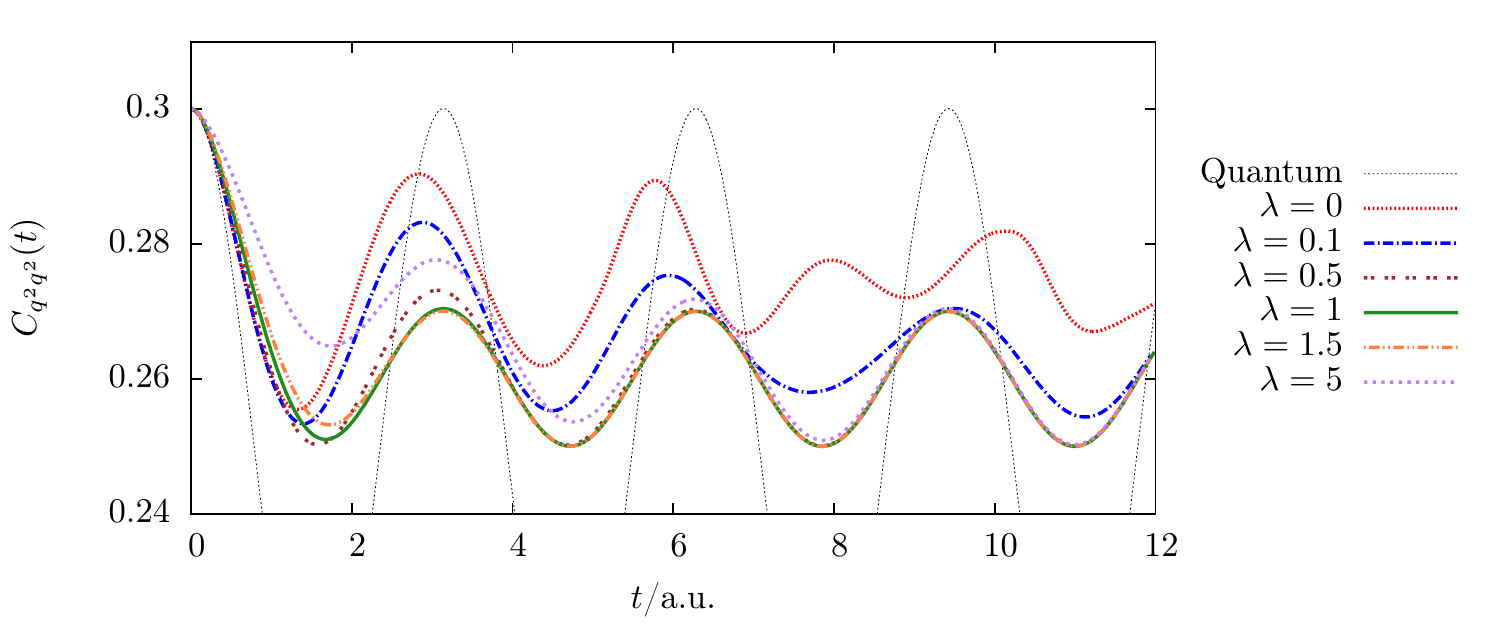}
 \caption{Position-squared autocorrelation function for a harmonic oscillator at $\beta=10$, showing the exact quantum result and TRPMD at a varying friction parameters. For clarity, the figure is zoomed in around the TRPMD correlation function.}
 \figl{lam}
\end{figure*}

We then examine the effect of different friction parameters in \figr{lam}, choosing the $\beta = 10$ system to exemplify the effect of damping. The $\lambda = 0$ (RPMD) result oscillates erratically, as in the third panel of \figr{qq}. Applying very small friction ($\lambda = 0.1$) noticeably improves the correlation function but contamination from higher normal modes is still evident. Around the optimal damping ($\lambda = 0.5$, $\lambda = 1$ and $\lambda = 1.5$) the correlation functions are extremely similar, settling to the correct frequency (though incorrect amplitude) after one oscillation. Increasing the friction yet further ($\lambda = 5$) causes the correct oscillation frequency (as all modes apart from the centroid are overdamped) but the slow decay of the heavily overdamped higher normal modes causes midpoint of the oscillation to decay slowly over time. These results (which can be derived analytically for the harmonic oscillator) are broadly consistent with those observed numerically in more complex systems (see Fig.~3 of Ref.~\onlinecite{ros14}), where a broad range of friction parameters around $0.5 \le \lambda \le 1.5$ led to similar results. 

For the position-squared autocorrelation function at least, it seems that TRPMD with $\lambda=1$ combines the best features of both RPMD and CMD; the correct zero-time value and the correct amplitude of oscillation, and that there exists a sizeable range of friction parameters around $\lambda = 1$ in which these qualitative features are captured. However, TRPMD with optimal friction is by no means perfect; the amplitude of oscillations decays within one oscillation and is far too small, since by this time all modes apart from the centroid are essentially completely damped. 


\subsection{Parabolic barrier}
\label{ssec:pb}
For an unbound, scattering system, or where barrier dynamics are required such as thermal rate calculation, we instead consider a parabolic barrier with a potential
\begin{align}
V(q) = -\tfrac{1}{2}m\omega_b^2.
\end{align}
We firstly observe that RPMD (and CMD) have qualitatively incorrect fluctuation dynamics at barriers; \cite{hel15rel} while all the Matsubara modes are scattering,
\begin{align}
 Q_j(t) = & Q_j\cosh(\omega_b t) + \frac{\bar P_j}{m\omega_b}\sinh(\omega_b t)\no\\
 &+ i\frac{\tilde \omega_j}{\omega}Q_{-j} \sinh(\omega_b t)
\end{align}
the RPMD higher normal modes are generally bound, with a frequency of $\bar \omega_j = \sqrt{\tilde \omega_j - \omega_b^2}$. As the temperature is lowered, modes become successively unbound, beginning with $j = \pm 1 $ at the `crossover' temperature \cite{ric09,hel13}
\begin{align}
 \beta_c = \frac{2\pi}{\hbar\omega_b} \eql{xover}
\end{align}
and the $j$th normal mode will become unbound when $\beta > |j|\beta_c$, but with a scattering (imaginary) frequency of $\sqrt{\omega_b - \tilde \omega_j^2}$. Despite these shortcomings, RPMD has been very accurate for rate calculation, partly because of the correct short-time TST limit \cite{hel15rel} (see Introduction), and since reaction above the crossover temperature is dominated by the motion of the centroid, which scatters at the correct imaginary frequency in RPMD, TRPMD and CMD. 

Considering the $j$th mode at a temperature $\beta < |j|\beta_c$ such that it is bound in RPMD ($\omega_b < |\tilde \omega_j|$), very weak friction ($\Gamma_{jj}/2 < \tilde \omega_j^2 - \omega_b^2$) leads to damped oscillatory motion as in \eqr{well}, but with $\acute \omega_j = \sqrt{-\omega_b^2 + \tilde \omega_j^2 - \Gamma_{jj}^2/4}$. Stronger friction leads to an overdamping solution,
\begin{align}
 \mathcal{Q}_j(t) = e^{-\Gamma_{jj}t/2} \Bigg[ & Q_j\cosh(\zeta_j t) \no\\
 & + \left(\frac{P_j}{m\zeta_j} + \frac{Q_j\Gamma_{jj}}{2\zeta_j}\right)\sinh(\zeta_j t)\Bigg] \eql{overdamp} 
\end{align}
where $\zeta_j = \sqrt{\omega_b^2 - \tilde\omega_j^2 + (\Gamma_{jj}/2)^2}$ can be considered the imaginary frequency counterpart to $\acute\omega_j$.

The presence of the $e^{-\Gamma_{jj}t/2}$ prefactor in \eqr{overdamp}, which in the oscillatory case of \eqr{well} causes damping but leaves the frequency untouched, means that no physical (i.e.\ real and positive \cite{mac08}) value of the friction parameter exists which would make \eqr{overdamp} have an unbound solution. Increasing friction merely causes the oscillator to become more overdamped.

If the normal mode is unbound in RPMD ($\beta > |j|\beta_c$ and therefore $\omega_b > |\tilde \omega_j|$) then \eqr{overdamp} still holds, but the solution has a scattering component for all $\Gamma_{jj}$ since $-\Gamma_{jj}/2 + \zeta_j > 0$. We would like to increase the escape rate from the RPMD value of $\sqrt{\omega_b^2 - \tilde\omega_j^2}$ to $\omega_b$, but adding friction only decreases the rate of escape from the barrier, as can be observed from the vanishing escape rate at high friction,
\begin{align}
 \lim_{\Gamma_{jj} \to \infty} -\Gamma_{jj}/2 + \zeta_j = 2\frac{\omega_b^2 - \omega_j^2}{\Gamma_{jj}}.
\end{align}
In order for the largest positive unbound solution (the highest root of the characteristic equation whose solution gives \eqr{overdamp}) to be the physical barrier frequency, the friction would have to be \emph{negative}; $\Gamma_{jj} = - \tilde\omega_j^2/\omega_b$.

For the (artificial) case of a reaction whose reaction co-ordinate is solely a single non-centroid normal mode, the above result is corroborated by Kramers theory \cite{kra40,nit06,hel15trpmd} which states that the transmission coefficient $\kappa(t)$ decreases with friction as 
\begin{align}
\ltti \kappa(t) \simeq \sqrt{1+\alpha^2} - \alpha\eql{klam}
\end{align}
where $\alpha = \Gamma_{jj}/2\bar \omega_j$ and $\bar\omega_j$ is the barrier frequency in ring-polymer space defined above. However, the transmission coefficient across a parabolic barrier is unity in Matsubara dynamics, and adding friction in TRPMD will only decrease this. Consequently, application of friction to RPMD will not ameliorate the qualitative problems with the RPMD higher normal modes at a barrier, and in some cases will worsen them. 

\section{Discussion}
\label{sec:dis}
The friction matrix $\Gamma_{jj'} = 2|\tilde \omega_j| \delta_{jj'}$ obtained section~\ref{ssec:hw} corresponds to critical damping of the ring polymer springs in the absence of an external potential, but not critical damping of the ring polymer modes in a harmonic oscillator (where the external frequency must also be considered), and can be determined without knowledge of the frequencies present in the external potential. Obviously, chemical systems will not be purely harmonic but in many systems (such as vibrating bond) this will be a reasonable approximation.

Previous literature has explored a range of scaled friction matrices of $\lambda \bm{\Gamma}$ and found $\lambda = 1/2$ to be optimal for some spectra, justifying this on the grounds of optimal sampling of the harmonic ring polymer potential energy \cite{ros14}, but also finding there to be a wide range of $\lambda$ close to $\lambda = 1/2$ in which results are broadly similar (as also seen in \figr{lam}). We suspect that a numerically favourable value of $\lambda = 1/2$ is due to interplay between shifting the frequencies of the higher normal modes to the external frequency (implying $\lambda=1$), moving the maximum in the spectral peak (implying $\lambda = 2^{-1/2}$), and avoiding harsh damping which would decorrelate the modes too quickly to capture their dynamics and broaden spectral peaks \cite{ros14com} (implying the weakest possible friction which removes spurious resonances). We can certainly find no reason to use $\lambda > 1$.

This definition of the friction matrix means that $\Gamma_{00} = 0$ (for all $\lambda$) meaning that the centroid is unthermostatted, so all the results which have previously been derived for TRPMD, such as its short-time error compared to the quantum result \cite{ros14}, still hold. A new result is that TRPMD, like RPMD, will have the exact Matsubara force on the centroid, since the error term does not act upon the centroid. Like RPMD and CMD but not LSC-IVR, the TRPMD dynamics will also satisfy detailed balance \cite{hel15trpmd}, and the error scaling in the higher normal modes in time will be the same as that for RPMD, namely $i\LIm - \mathcal{A}_{\rm wn}^{[M]\dag}\propto 1/\beta\hbar$. 

This choice of friction matrix also means that the TRPMD correlation function of a linear operator will deviate from the Matsubara correlation due to higher-order coupling between the centroid dynamics and the damping (and random kicks) of the higher normal modes via anharmonicity in the potential. This causes slight broadening of spectral lines (a far smaller issue than the curvature problem of CMD or the spurious resonances of RPMD) \cite{ros14}, but the extra friction noticeably slows reaction rates beneath the crossover temperature \cite{hel15trpmd} where the unbound and thermostatted higher normal modes are part of the optimal dividing surface \cite{ric09}. For nonlinear operators, TRPMD (like RPMD) would be expected to break down faster than for linear operators due to the error term only acting directly on the higher normal modes, though the example of the position-squared autocorrelation function given above suggests that with a careful choice of friction the breakdown may not be too drastic. 


Although the analysis for a parabolic barrier in section~\ref{ssec:pb} does not suggest that the TRPMD rate will ever be closer to the Matsubara (and therefore quantum) rate, TRPMD could be computationally advisable above the crossover temperature (where passage over the barrier is dominated by motion of the unthermostatted centroid) since the TRPMD trajectories may sample the path-integral phase space more efficiently than the RPMD trajectories \cite{hel15trpmd}, and same may be true for other observable properties which are dominated by barrier crossing, such as diffusion \cite{ros14}\footnote{Rossi and Manolopoulos, private communication, (2015).}. The TRPMD time-evolution is also simpler computationally since the same dynamics can be used for thermostatting and computation of the correlation function \cite{ros14}. Nevertheless, these results suggest that beneath the crossover temperature, TRPMD is not to be advised for reaction rates, a result broadly supported by numerical tests in one-dimensional and multidimensional gas-phase systems \cite{hel15trpmd}. 

All the results presented here generalize immediately to multidimensional systems, where the friction is applied in $F(N-1)$ normal modes and springs exist between $N$ replicas of the physical system. For nonlinear operators one cannot, in general, easily relate the Kubo and Generalized Kubo forms [\eqsr{kub}{genk}] (the position-squared operator explored above being an exception). For reaction rates involving the highly nonlinear flux and side operators this is resolved by relating the generalized Kubo form to the exact quantum expression when there is no recrossing of the path-integral dividing surface (and those orthogonal to it in path-integral space) by the exact quantum dynamics of the system \cite{hel13,alt13,hel13unique}, such that TRPMD rate theory will give the exact quantum rate in the absence of recrossing by either the TRPMD dynamics or exact quantum dynamics. \cite{hel15trpmd}

\section{Conclusions}
\label{sec:con}
In this article we have shown, for the first time, how to obtain thermostatted ring polymer molecular dynamics (TRPMD) from exact quantum dynamics by a series of approximations, each with an analytic error term. We firstly discard fluctuations of the highest $N-M$ normal modes from the exact quantum time evolution, giving Matsubara dynamics \cite{hel15}. To derive a computationally tractable approximation to Matsubara dynamics, we perform a contour integral in the momenta (where we assume the edge terms to be zero), giving a correlation function with the (real) ring polymer distribution, but whose Liouvillian is complex. We then replace the imaginary part of the complex Liouvillian with a white-noise Fokker-Planck term, giving TRPMD.

Each of these approximations has its limitations and benefits. The primary consequence of discarding the fluctuations of the highest normal modes from the exact quantum dynamics (leading to Matsubara dynamics) is neglect of interference effects and mixing of quantum states. In physical systems this is seen as the failure of TRPMD to replicate the Fermi resonance in the Zundel cation \cite{ros14} (CMD and RPMD also fail here \cite{ros14}, as would be expected as they too are approximations to Matsubara dynamics \cite{hel15rel}). However, discarding these fluctuations leads to a classical-like dynamics which preserves the quantum Boltzmann distribution. \cite{hel15}

We then show that a careful choice of the friction matrix (which is system independent and known in advance) will cause all ring polymer normal modes to oscillate at the correct frequency in a harmonic potential, and therefore will reproduce the correct frequency of oscillation of the position-squared autocorrelation function and the correct $t=0$ value; neither CMD nor RPMD will replicate both of these properties. However, the oscillations' amplitude is too small, and we suspect that a generalized Langevin equation \cite{cer10,nit06} may be more successful than a simple white noise thermostat, in that it may be constructed to produce the correct frequency of oscillation of the higher normal modes but with smaller damping (and maybe even the correct maxima in the position and momentum autocorrelation spectra)\footnote{Michele Ceriotti, private communication, 2015.}. The same analysis, but for a parabolic barrier, shows that no physical friction parameter will solve the qualitative innacurracies in the higher ring polymer normal mode fluctuations. Usage of unphysical negative friction \cite{mac08,che02} as a possible solution to this problem is left as further work. Future research could also include extension to non-adiabatic systems where RPMD has been successful. \cite{hel11,ana13,ana10,men14,ric13,ric14,men11,kre13}

In closing, the results presented here give an \emph{a priori} prescription for when to use TRPMD: it should be used for computation of spectra and other properties of bound systems where the correct oscillation frequencies are required, and avoided for rate calculation beneath the crossover temperature.

\appendix
\section{Matsubara modes}
\label{ap:mat}
The ring-polymer normal modes are defined as
\begin{align}
 Q_j = \smiNz \frac{T_{ij}}{\sqrt{N}} q_i 
\end{align}
where $j = -N/2+1, \ldots,0,\ldots, N/2$ and likewise for $\bP$, where
\begin{align}
 T_{ij} = 
 \left\{
 \begin{array}{ll}
  N^{-1/2} & j=0 \\
  \sqrt{2/N} \sin(2\pi ij/N) & 1 \leq j \leq N/2 - 1 \\
  N^{-1/2} (-1)^i & j = N/2 \\
  \sqrt{2/N} \cos(2\pi ij/N) & -N/2 + 1 \leq j \leq - 1
 \end{array}
 \right.
\end{align}
where the $j=N/2$ mode is omitted if $N$ is odd.\footnote{For mathematical simplicity we consider odd $N$ here, even $N$ leads to the same result but with more algebra{\protect  \cite{hel15}}.} The transformation is not unitary, but defined such that the normal modes converge in the \largeN\ limit. This leads to frequencies in the complex Boltzmann distribution of
\begin{align}
 \omega_j = \frac{2 \sin(j\pi/N)}{\betaN\hbar} \eql{rpfreq}
\end{align}
which, for large $N$ and finite $j$, become the Matsubara frequencies \cite{mat55}
\begin{align}
 \tilde \omega_j = \lNti \omega_j = \frac{2\pi j}{\beta \hbar}.
\end{align}
The observables $A(\bQ)$ and $B(\bQ)$ are obtained by making by substituting
\begin{align}
 q_i = \smjMM T_{ij} \sqrt{N} Q_{j}
\end{align}
into $A(\bq)$ and $B(\bq)$ respectively, which also leads to a `Matsubara potential',
\begin{align}
 \UMQ = \frac{1}{N}\smiNz V\!\left(\smjMM T_{ij} \sqrt{N} Q_j \right).
\end{align}

\section{Equivalence of quantum and Matsubara correlation functions}
\label{ap:qm}
To show that the Matsubara correlation function is equivalent to \eqr{qqqm}, we firstly calculate the Matsubara correlation function using the harmonic analysis in the supplementary material of Ref.~\onlinecite{hel15rel}, giving
\begin{widetext}
\begin{align}
  C_{q^2q^2}^{[N]}(t) = &\frac{1}{\beta^2m^2\omega_h^4}\Bigg[\cos(2\omega_h t)\smjMM \frac{1}{1 + (\tilde \omega_j/\omega_h)^2} + \smjMM \frac{1-(\tilde\omega_j/\omega_h)^2}{(1 + (\tilde \omega_j^2/\omega_h))^2}  \no \\
  & + \smjMM\smkMM\frac{1}{1 + (\tilde \omega_j/\omega_h)^2}\frac{1}{1 + (\tilde \omega_k/\omega_h)^2} \Bigg]\eql{sqterm}
\end{align}
\end{widetext}
The Matsubara frequency summation \cite{alt10} is performed by examining the integral
\begin{align}
 \oint dz \frac{\cot(z)}{z^2 + x^2} \eql{contint}
\end{align}
around a circle of infinite radius, origin zero, we find
\begin{align}
 x\coth(x) = \lim_{M\to\infty} \smjMM \frac{1}{1+ (j\pi/x)^2} \eql{decep}
\end{align}
and by differentiation of \eqr{decep}, that
\begin{align}
 x^2[\coth^2(x) - 1] = & \lim_{M\to\infty} \smjMM \frac{1 - (j\pi/x)^2}{(1+(j\pi/x)^2)^2} \eql{ddecep}.
\end{align}
Subsituting $x = \beta\hbar\omega_h/2$ into \eqr{decep} and \eqr{ddecep}, and these expressions into \eqr{sqterm} gives \eqr{qqqm} as required. 

\section{Analyticity in the complex plane}
\label{ap:contour}
Consider an observable $B(\bP,\bQ,t)$, which is propagated by the Liouvillian
\begin{align}
 \mathcal{L} = \left(\nabla_\bP H\right)\cdot \nabla_\bQ - \left(\nabla_\bQ H\right)\cdot \nabla_\bP 
\end{align}
where $H$ is the Hamiltonian of the system and an analytic, but not necessarily real, function of $\bP$ and $\bQ$.
The propagation is formally
\begin{align}
 \dd{}{t}B(\bP,\bQ,t) = & \mathcal{L} B(\bP,\bQ,t) \\
 B(\bP,\bQ,t) = & e^{\mathcal{L}t}B(\bP,\bQ,0) \eql{exp}
\end{align}
This (obviously) requires $B(\bP,\bQ,t)$ to be single valued, and the exponentiated expression \eqr{exp} to exist. If $B(\bP,\bQ,t)$ is an analytic function for all values of $\bz$, then (by the Cauchy-Riemann relations)
\begin{align}
 \ddp{}{P_j^*}B(\bP,\bQ,t) = 0 \ \forall j
\end{align}
where $P_j^*$ is the complex conjugate of $P_j$ (and likewise for $Q_j^*$). If $H$ is analytic then
\begin{align}
  \ddp{}{P_j^*} H = 0\ \forall j
\end{align}
which means that (using $H$ being continuous, Schwarz' theorem and therefore $\ddp{}{P_j^*}\ddp{}{P_j} = \ddp{}{P_j}\ddp{}{P_j^*}$) the commutation relations exist
\begin{align}
\ddp{}{P_j^*}\mathcal{L} = & \mathcal{L}\ddp{}{P_j^*}
\end{align}
Using the definition of an exponential as its power expansion we then see, 
\begin{align}
 \ddp{}{P_j^*} B(\bP,\bQ,t) = &\ddp{}{P_j^*} e^{\mathcal{L}t}B(\bP,\bQ,0) \\
 = & e^{\mathcal{L}t}\ddp{}{P_j^*}B(\bP,\bQ,0) \\
 = & 0
\end{align}
so $B(\bP,\bQ,t)$ remains an analytic function of $P_j$ for all time (and likewise for $Q_j$). This is true $\forall j$ (and $\forall t$), and by Hartog's Theorem, true for $B(\bP,\bQ,t)$ everywhere. This means that $B(\bP,\bQ,t)$ obeys the Cauchy Riemann relations and can have no poles in the complex plane. The Boltzmann distribution is also holomorphic, and provided that the zero-time observable $A(\bP,\bQ,0)$ is also holomorphic (which almost all physical observables are) the entire integrand of \eqr{mat} will be.

We then complete the square in the complex Matsubara distribution, giving \eqr{mat} where the edges of the rectangle used in the contour integration are
\begin{align}
 \mathcal{E}(t) = & \lim_{\bm{\pi} \to \infty} \int d\bQ \left[\pjMM i\int_0^{m\omega_jQ_{-j}} d\Pi_j\right] \no\\
 & \qquad \times e^{-\beta [ H(\bm{\pi} + i \bm{\Pi}, \bQ) - i \theta(\bm{\pi} + i \bm{\Pi}, \bQ)]} A( \bQ) e^{\mathcal{L}^{[M]}_{\bm{\pi} + i\bm{\Pi}} t} B(\bQ) \no\\
 & + \lim_{\bm{\pi} \to -\infty} \int d\bQ \left[\pjMM i\int_0^{m\omega_jQ_{-j}} d\Pi_j\right] \no\\
 & \qquad \times e^{-\beta [ H(\bm{\pi} + i \bm{\Pi}, \bQ) - i \theta(\bm{\pi} + i \bm{\Pi}, \bQ)]} A( \bQ) e^{\mathcal{L}^{[M]}_{\bm{\pi} + i\bm{\Pi}}t} B(\bQ)\eql{edge}
\end{align}
where $\pi_j = \Re P_j$, $\Pi_j = \Im P_j$, and $\mathcal{L}^{[M]}_{\bm{\pi} + i\bm{\Pi}}$ is the Matsubara Liouvillian \eqr{lmat} continued into the complex plane.

The edge terms can be proven to be zero in a number of limits. Specifically, for $A(\bQ)$ and $B(\bQ)$ which are at most exponential in $\bP$ and/or $\bQ$, the edge terms will vanish when the trajectories are real ($\bm{\Pi} = 0$) where conservation of energy arguments can be used in a bound system and in a scattering system whose potential tends to a constant value far out. The edges will also be zero in any system at $t=0$ where the momentum integral can be evaluated analytically, and where discarding $\mathcal{L}^{[M]}_{\Im}$ (and thereby keeping the trajectories real) is no approximation, namely up to $\mathcal{O}(t^2)$ for nonlinear operators and $O(t^6)$ for linear operators \cite{hel15rel,ros14}.

For systems where the trajectories are known analytically, such as a free particle, parabolic well and barrier, even though $\bm{\pi}(t) \to \infty$ as $\bm{\pi}(0) \to \infty$, careful consideration of the limits and application of l'H\^opital's rule shows that the edge term still vanishes.

Despite the above promising results, trajectories in the complex plane are frequently not bounded \cite{ben08} and in general it is difficult to determine whether or not terms of the form in \eqr{edge} will converge \cite{aar10trust} for any general potential. A proof of whether $\mathcal{E}(t)$ can be neglected in any general case is left as further work.

\section{TRPMD position-squared correlation functions}
The correlation function can be evaluated by considering each normal mode separately and deriving the correlation function for a single harmonic oscillator \cite{cer10phd,nit06,gar09}. For $0 \leq \lambda \leq 1$, the correlation function is
\begin{widetext}
\begin{align}
 C_{q^2q^2}^{[N]}(t) = \frac{1}{\beta^2m^2}\smjMM \frac{1}{\omega_h^2 + \tilde \omega_j^2} \left\{\frac{2 e^{-\Gamma_{jj} t}}{\omega_h^2 + \tilde \omega_j^2} 
 \left[\cos(\acute \omega_j t) + \frac{\gamma_j}{2 \acute \omega_j} \sin(\acute \omega_j t)\right]^2 + \smkMM \frac{1}{\omega_h^2 + \tilde \omega_k^2}\right\} \eql{q2t}
\end{align}
with $\acute \omega_j$ defined in \eqr{coj}. If $\lambda > 1$, we define
\begin{align}
 j_{\rm cut} = \left\lfloor  \frac{\beta\hbar\omega_h}{2\pi \sqrt{\lambda^2-1}} \right\rfloor
\end{align}
where $\lfloor \cdot \rfloor$ is the floor function, so all modes with $|j| > j_{\rm cut}$ will be overdamped. The correlation function is
\begin{align}
 C_{q^2q^2}^{[N]}(t) = & 
 \frac{1}{\beta^2m^2}\sum_{j=-j_{\rm cut}}^{j_{\rm cut}} \frac{1}{\omega_h^2 + \tilde \omega_j^2} \Bigg\{\frac{ e^{-\Gamma_{jj} t}}{\omega_h^2 + \tilde \omega_j^2}
  \left[1 + \frac{\Gamma_{jj}^2}{4\zeta_j^2} + \left(1 - \frac{\Gamma_{jj}^2}{4\zeta_j^2}\right)\cos(2\zeta_j t) + \frac{\Gamma_{jj}}{\zeta_j}\sin(2\zeta_jt) \right] + \smkMM \frac{1}{\omega_h^2 + \tilde \omega_k^2}\Bigg\} \no\\
 & + \frac{2}{\beta^2m^2}\sum_{j=j_{\rm cut}+1}^{M/2-1} \frac{1}{\omega_h^2 + \tilde \omega_j^2} \Bigg\{\frac{ e^{-\Gamma_{jj} t}}{\omega_h^2 + \tilde \omega_j^2}
  \left[1 - \frac{\Gamma_{jj}^2}{4\zeta_j^2} + \left(1 + \frac{\Gamma_{jj}^2}{4\zeta_j^2}\right)\cosh(2\zeta_j t) + \frac{\Gamma_{jj}}{\zeta_j}\sinh(2\zeta_jt) \right]
 + \smkMM \frac{1}{\omega_h^2 + \tilde \omega_k^2}\Bigg\}
\end{align}
\end{widetext}
where we have noted that contributions from modes $j$ and $-j$ are the same. If a mode is critically damped then the term in square brackets for that mode becomes $2 + \Gamma_{jj}^2t^2/2 + 2\Gamma_{jj}t$.

\section*{Acknowledgements}
This work was supported by a Research Fellowship from Jesus College, Cambridge. The author wishes to thank Stuart Patching for suggesting the contour integral in \eqr{contint}, and is also grateful for corrections to and comments on the manuscript from Michael Willatt, advice from Stuart Althorpe and Michele Ceriotti, and for helpful discussions with Robert Whelan and Adam Harper.



\end{document}